\documentclass[preprint,authoryear,review,12pt]{elsarticle}

\usepackage{graphics}
\usepackage{graphicx}
\usepackage{subfigure}

\usepackage{amssymb}
\usepackage{amsthm}
\usepackage{amsmath}

\begin{document}

\begin{frontmatter}



\title{Robust Bias Estimation for Kaplan--Meier Survival Estimator with Jackknifing}


\author[mhr]{Md Hasinur Rahaman Khan}
\address[mhr]{Institute of Statistical Research and Training, University of Dhaka, Bangladesh}

\author[jehs]{J. Ewart H. Shaw}
\address[jehs]{Department of Statistics, University of Warwick, UK}

\begin{abstract}
For studying or reducing the bias of functionals of the Kaplan--Meier survival estimator,
the jackknifing approach of Stute and Wang (1994) is natural.
We have studied the behavior of the jackknife estimate of bias under different
configurations of the censoring level,
sample size, and the censoring and survival time distributions.
The empirical research reveals some new findings about robust calculation
of the bias, particularly for higher censoring levels.
We have extended their jackknifing approach to cover the case where the largest observation is censored, using the imputation methods for the largest observations proposed in Khan and Shaw (2013b\nocite{Kha:sha:13:OnDeal}).
This modification to the existing formula reduces the number of conditions for creating jackknife bias estimates to one from the original two,
and also avoids the problem that the Kaplan--Meier estimator can be badly underestimated by the existing jackknife formula.
\end{abstract}


\end{frontmatter}

\section{Introduction}\label{sec:km effect}
Suppose that there is a random sample of $n$ individuals. Let $T_i$ and $C_i$ be the random variables
that represent the lifetime and censoring time for the $i$th individual.
We also assume $T_i$ has unknown distribution function $F$. The Kaplan-Meier (K--M)
estimator, $\hat{F}^{KM}$ (Kaplan and Meier, 1958\nocite{kapl:meie:58:nonpara}) is then defined by
\begin{equation}
1-\hat{F}^{KM}(t)=\prod_{y_{(i)}\le y}\Big(\frac{n-i}{n-i+1}\Big)^{\delta_{(i)}},
\label{eq:km}
\end{equation}
where $Y_{(1)}\le \cdots \le Y_{(n)}$ are the ordered observations
(censored and uncensored lifetimes),
$\delta_{(i)}=1$ if $Y_{(i)}$ is observed and
$\delta_{(i)}=0$ if $Y_{(i)}$ is censored,
ties between censoring times are treated as if the former precede the latter,
and other ties are ordered arbitrarily.
Suppose that $S$ is a given statistical function so
that $S(F)$ is the parameter of interest.
It follows from Stute (1994\nocite{stut:94:thebias})
that if $S$ is nonlinear then the K--M based estimator, $S(F^{KM})$,
is biased.
Stute (1994\nocite{stut:94:thebias})
also discussed the situation where the bias arises even for linear $S$
when the data of interest are partially observable.
Now for any $F$-integrable function
$\varphi$, the corresponding estimator of the parameter of interest,
$S(\hat{F}^{KM})$ is defined by the K--M integral $\int
\varphi(Y_{(i)})\,\mbox{d}\hat{F}^{KM}$.

The K--M estimator is well known to be unbiased if there
is no random censorship but it becomes biased under censorship. Gill
(1980\nocite{gill:80:censoring}) was the first to bound the bias of
$\hat{F}^{KM}:-F\,H\leq E(\hat{F}^{KM})-F\leq 0$, where $H$ is the
distribution function of $Y$.
Mauro (1985\nocite{mau:85:acom})
extended this result to arbitrary K--M integrals with non-negative
integrands. Zhou (1988\nocite{zhou:88:twosided}) proved that the
bias of the K--M estimator functional decreases at an exponential
rate, and always underestimates the true value. He established the
lower bound: $-\int \varphi\, H\, F(dt)\leq\mbox{bias}(\int
\varphi\,\mbox{d}\hat{F}^{KM})\leq 0$. Stute
(1994\nocite{stut:94:thebias}) derived the exact formula for the
bias of $\int \varphi\,\mbox{d}\hat{F}^{KM}$ for a general
Borel-measurable function, $\varphi$. He also discussed the effect
of light, medium or heavy censoring on the bias of $\int
\varphi\,\mbox{d}\hat{F}^{KM}$.
Stute and Wang
(1994\nocite{stut:94b:thejack}) derived an explicit formula for the
jackknife estimate of the bias of $\int
\varphi(Y_{(i)})\,\mbox{d}\hat{F}^{KM}$.
They also showed that
jackknifing can lead to a considerable reduction of the bias. Four years later, Shen (1998\nocite{Shen:98:problem}) proposed another explicit formula for jackknife estimate of bias of $\int\varphi(T_{(i)}^*)\,\mbox{d}\hat{F}^{KM}$. He used delete-2 jackknifing where two observations are deleted. It follows from Shen (1998\nocite{Shen:98:problem}) that the formula based on delete-2 doesn't show any further improvement on the delete-1 formula. Stute (1996\nocite{stut:96:thejack_var}) also proposed a jackknife estimate of the variance of $\int\varphi(Y_{(i)})\,\mbox{d}\hat{F}^{KM}$.

As mentioned in Stute and Wang (1994\nocite{stut:94b:thejack}),
under random censorship the estimator $S(\hat{F}^{KM})$ becomes the K--M integral
\begin{equation}
S(\hat{F}^{KM})=\sum_{i=1}^{n}w_i\,\varphi(Y_{(i)})\equiv \hat{S}^{KM}_{\varphi},\qquad i=1,\cdots,n
\label{eq:kmInt}
\end{equation}
where the the K--M weights $w_i$ are the sizes of the
jumps by which the K--M estimator of $F$ changes at the uncensored
points $Y_{(i)}$, given by
\begin{equation}
w_{1}=\frac{\delta_{(1)}}{n}, \qquad
 w_{i}=\frac{\delta_{(i)}}{n-i+1}\prod_{j=1}^{i-1}\Big(\frac{n-j}{n-j+1}\Big)^{\delta_{(j)}},\quad i=2,\cdots,n.
\label{eq:kmweights}
\end{equation}
A detailed study of the $w_i$'s in connection with
the strong law of large numbers under censoring has been carried out
in Stute and Wang (1993\nocite{stut:wang:93:thestr}).

The jackknife estimate of bias for the K--M integral (Eq.~\ref{eq:kmInt}) is given by
\begin{equation}
\mbox{Bias}\,(\hat{S}^{KM}_{\varphi})=-\frac{n-1}{n}\,\varphi(Y_{(n)})\,\delta_{(n)}\,(1-\delta_{(n-1)})\prod_{j=1}^{n-2}\Big(\frac{n-1-j}{n-j}\Big)^{\delta_{(j)}}.
\label{eq:jkbias}
\end{equation}
The associated bias corrected jackknife estimator is therefore given by
\begin{equation}
\tilde{S}^{KM}_{\varphi}=\hat{S}^{KM}_{\varphi}-\mbox{Bias}\,(\hat{S}^{KM}_{\varphi}).
\label{eq:jkest}
\end{equation}

\section{Modified Jackknife Bias for K--M Lifetime Estimator}
When no censoring is present, $\hat{F}^{KM}$ reduces to
the usual sample distribution estimator $\hat{F}$ that assign
weight $\frac{1}{n}$ to each observation.
With censoring, the weighting method (\ref{eq:kmweights}) gives zero
weight to the censored observations $Y_{(.)}^+$,
causing particular problems if the largest datum is censored (i.e.~$\delta_{(n)}=0$).
As a first step one may apply Efron's (1967\nocite{efro:67:the})
tail correction approach: reclassify $\delta_{(n)}=0$ as $\delta_{(n)}=1$.
In order to reduce estimation bias and inefficiency,
Khan and Shaw (2013b\nocite{Kha:sha:13:OnDeal})
proposed five alternatives to Efron's approach,
that can lead to more efficient and less biased estimates. The approaches are summarised in Table~\ref{table:imp_methods}.
\begin{table}[h]
\centering
\caption{The imputation approaches from Khan and Shaw (2013b).}
\begin{tabular}{l}\hline
$W_{\tau_{m}}$: Adding the Conditional Mean\\
$W_{\tau_{md}}$: Adding the Conditional Median\\
$W_{\tau^{\ast}_{m}}$: Adding the Resampling-based Conditional Mean\\
$W_{\tau^{\ast}_{md}}$: Adding the Resampling-based Conditional Median\\
$W_{\nu}$: Adding the Predicted Difference Quantity\\\hline
\end{tabular}
\label{table:imp_methods}
\end{table}
The first four approaches are based on the underlying regression assumption relating lifetimes and covariates
(e.g.,~the AFT model), and the fifth approach $W_{\nu}$, is based on only the random
censorship assumption.

The jackknife bias in Eq.~(\ref{eq:jkbias}) is non-zero
if and only if the largest datum is uncensored, $\delta_{(n)}=1$,
and the second largest datum is censored, $\delta_{(n-1)}=0$.
Stute and Wang (1994\nocite{stut:94b:thejack}) state that if $\delta_{(n)}=0$,
then the corresponding observation doesn't contain enough information
about $F$ to make a change of $\hat{S}^{KM}_{\varphi}$ desirable.
This inability to estimate bias if $\delta_{(n)}=0$
is a major limitation of the jackknife bias formula.

If ($\delta_{(n-1)}=0,~\delta_{(n)}=0$),
then we can obtain a modified jackknife estimate of bias by imputing the largest datum,
for example using any of the approaches given in Table~\ref{table:imp_methods}.
From Eq.~(\ref{eq:kmInt})
this gives the modified estimator
\begin{eqnarray}
\hat{S^*_{\varphi}}^{KM}\equiv \sum_{i=1}^{n-1}w_i\,\varphi(Y_{(i)})+\acute{w}_n\,\varphi(\tilde{Y}_{(n)}),\qquad i=1,\cdots,n-1,
\label{eq:modKME}
\end{eqnarray}
where $\tilde{Y}_{(n)}$ is the imputed largest observation,
and $\acute{w}_n$ is the corresponding adjusted K--M weight
\[
\acute{w}_n=w_n+\frac{n-1}{n}\,\prod_{j=1}^{n-2}\Big(\frac{n-1-j}{n-j}\Big)^{\delta_{(j)}}
\]
as suggested in Stute and Wang (1994\nocite{stut:94b:thejack}) for the pair $(\delta_{(n-1)}=0,~\delta_{(n)}=1)$.
The modified estimator (\ref{eq:modKME}) is also obtained when imputing in the situation $(\delta_{(n-1)}=1,~\delta_{(n)}=0)$.
In this case the K--M weight to $\tilde{Y}_{(n)}$ is not adjusted and we arrive at the estimator
\begin{eqnarray*}
\hat{S^*_{\varphi}}^{KM}\equiv \sum_{i=1}^{n-1}w_i\,\varphi(Y_{(i)})+w_n\,\varphi(\tilde{Y}_{(n)}),\qquad i=1,\cdots,n-1.
\label{eq:modKME2}
\end{eqnarray*}

So unlike the actual jackknife formula the modified approach doesn't impose any condition on the censoring status of $Y_{(i)}$. The modified estimate of bias is given by
\begin{equation}
\mbox{Bias}\,(\hat{S^*_{\varphi}}^{KM})=-\frac{n-1}{n}\,\varphi(\tilde{Y}_{(n)})\,\delta^*_{(n)}\,(1-\delta_{(n-1)})\prod_{j=1}^{n-2}\Big(\frac{n-1-j}{n-j}\Big)^{\delta_{(j)}},
\label{eq:jkbias*}
\end{equation}
where $\delta^*_{(n)}$ is the modified censoring indicator for $\tilde{Y}_{(n)}$.
With the above approach, $\delta^*_{(n)}$ is always 1.
It follows from Eq.~(\ref{eq:jkbias*}) the larger bias quantity because $\tilde{Y}_{(n)}>Y_{(n)}$.
The modified bias corrected jackknife estimator is then defined by
\begin{equation}
\tilde{S^*_{\varphi}}^{KM}=\hat{S^*_{\varphi}}^{KM}-\mbox{Bias}\,(\hat{S^*_{\varphi}}^{KM}).
\label{eq:jkest*}
\end{equation}

The K--M estimates under both approaches for the four pairs are summarized in Table~\ref{table:km.summary}.
\begin{table}[h]
\caption{K--M lifetime estimates by censoring indicators for the last two observations.}
\centering
\begin{tabular}{lcc}\hline
K--M estimate &$\delta_{(n-1)}$&$\delta_{(n)}$\\\hline
$\hat{S^*_{\varphi}}^{KM}+\frac{n-1}{n}\,\varphi(\tilde{Y}_{(n)})\,\delta^*_{(n)}\,(1-\delta_{(n-1)})\prod_{j=1}^{n-2}\Big(\frac{n-1-j}{n-j}\Big)^{\delta_{(j)}}$&$0$&$0$\\
$\hat{S^*_{\varphi}}^{KM}$&$1$&$0$\\
$\hat{S}^{KM}_{\varphi}$&$1$&$1$\\
$\hat{S}^{KM}_{\varphi}+\frac{n-1}{n}\,\varphi(\tilde{Y}_{(n)})\,\delta_{(n)}\,(1-\delta_{(n-1)})\prod_{j=1}^{n-2}\Big(\frac{n-1-j}{n-j}\Big)^{\delta_{(j)}}$&$0$&$1$\\
\hline
\end{tabular}
\label{table:km.summary}
\end{table}

We investigate below the effect of censoring on the K--M
estimator $S(\hat{F}^{KM})$ based on both the actual and the modified
jackknife bias formula.
For computational simplicity we look only at the K--M mean lifetime estimator,
obtained by replacing $\varphi(y)$ by $y$ in Eq.~(\ref{eq:kmInt}).
Note that researchers in reliability are very often interested
in estimating the mean lifetime of a component,
and that the K--M mean lifetime estimate also has an important role in
Health Economics, for example, in a ``QTWIST'' analysis (Glasziou et
al.~1990\nocite{Glas:sime:gelb:90:quali}).
Obviously the behaviour of
the K--M mean lifetime estimator depends on the nature of the
distribution being estimated and the degree of censoring, although
the true distribution of censored data is generally unknown.
We therefore conducted simulation studies
to demonstrate the behavior of the K--M mean lifetime estimator in the
presence of right censoring.
We assume that the lifetimes and censoring times have independent distributions.

Note that the mean survival time can be defined as the area under the survival
curve, $S(t)$ (Kaplan and Meier, 1958\nocite{kapl:meie:58:nonpara}).
A nonparametric estimate of the mean survival time can also be
obtained by substituting the K--M mean estimator for the unknown
survival function
$\hat{\mu}=\int_{0}^{\infty}\,\hat{S}(t)\,\mbox{d}t$.
Stute (1994\nocite{stut:94b:thejack}) proposed a bias
corrected jackknife estimator for the K--M mean lifetime.
When the observations are subject to right censoring, the usual mean estimator of the mean lifetime is not
appropriate (Datta, 2005\nocite{datt:05:estimati}).
The reason is that the censoring leads to an inconsistent estimator
that underestimates the true mean and
the bias worsens as the censoring increases.

\section{Simulation Study}
This section reports on three simulation based examples.
The first example extends the Koziol-Green model simulations of Stute and Wang (1994\nocite{stut:94b:thejack}).
The second example considers various skewed distributions for survival times
and corresponding distributions for the associated censored times.
The third example uses a log-normal AFT model
where the event times are assumed to be associated with several covariates.
\subsection{Koziol-Green Model based Example}
This extends the simulations of the Koziol-Green proportional hazards
model from Stute and Wang
(1994\nocite{stut:94b:thejack}).
Under this model both $T$ and $C$
were exponentially distributed: $T\sim \mbox{Exp}\,(1)$ and $C\sim \mbox{Exp}\,(\lambda)$,
with varying $\lambda$'s.
Four different sample sizes $n = 30,\,50,\,100,\,150$ are used.
For each sample,
$100,000$ simulation runs are drawn and the bias and variance of both the
mean lifetime estimators $\hat{S}^{KM}_{\mbox{mean}}$ and
$\tilde{S}^{KM}_{\mbox{mean}}$ are computed. The bias and its variance
are shown in Table \ref{tab:KMbias-var.KGM} and \ref{tab:KM.var}
(the first sub-table for both tables) respectively.

\begin{table}[h]\centering
\caption{Simulation results based on the Koziol-Green model for
the bias of the four K--M mean lifetime estimators
$\hat{S}^{KM}_{\mbox{mean}}$, $\tilde{S}^{KM}_{\mbox{mean}}$, $\hat{S^*}^{KM}_{\mbox{mean}}$ and $\tilde{S^*}^{KM}_{\mbox{mean}}$.} \scalebox{0.7}{
\begin{tabular}{lcccccccccc}\hline
$P_{\%}$&n=30&n=50&n=100&n=150&&n=30&n=50&n=100&n=150\\\hline
&\multicolumn{4}{c}{Bias of $\hat{S}^{KM}_{\mbox{mean}}$}&&\multicolumn{4}{c}{Bias of $\tilde{S}^{KM}_{\mbox{mean}}$}\\
10& -0.155& -0.114& -0.073& -0.055&& -0.154& -0.114& -0.073& -0.056\\
20& -0.197& -0.157& -0.107& -0.085&& -0.191& -0.155& -0.107& -0.086\\
30& -0.250& -0.205& -0.151& -0.126&& -0.233& -0.195& -0.146& -0.123\\
40& -0.304& -0.265& -0.209& -0.178&& -0.267& -0.239& -0.193& -0.164\\
50& -0.364& -0.327& -0.278& -0.248&& -0.295& -0.268& -0.237& -0.215\\
60& -0.409& -0.389& -0.349& -0.328&& -0.287& -0.281& -0.263& -0.255\\
70& -0.430& -0.426& -0.413& -0.396&& -0.224& -0.234& -0.246& -0.245\\
80& -0.402& -0.417& -0.428& -0.428&& -0.082& -0.097& -0.127& -0.141\\
90& -0.280& -0.304& -0.335& -0.346&&  0.161&  0.178&  0.171&  0.164\\
&\multicolumn{4}{c}{Bias of $\hat{S^*}^{KM}_{\mbox{mean}}$}&&\multicolumn{4}{c}{Bias of $\tilde{S^*}^{KM}_{\mbox{mean}}$}\\
10& -0.208& -0.147& -0.090& -0.067&& -0.207& -0.147& -0.090& -0.068\\
20& -0.259& -0.202& -0.132& -0.104&& -0.252& -0.200& -0.132& -0.104\\
30& -0.326& -0.261& -0.186& -0.155&& -0.309& -0.251& -0.181& -0.152\\
40& -0.391& -0.335& -0.260& -0.218&& -0.354& -0.310& -0.243& -0.205\\
50& -0.465& -0.407& -0.343& -0.304&& -0.396& -0.349& -0.303& -0.271\\
60& -0.511& -0.481& -0.426& -0.400&& -0.389& -0.372& -0.341& -0.327\\
70& -0.518& -0.512& -0.495& -0.475&& -0.312& -0.320& -0.328& -0.325\\
80& -0.463& -0.481& -0.496& -0.498&& -0.162& -0.162& -0.195& -0.210\\
90& -0.304& -0.331& -0.367& -0.380&&  0.151&  0.151&  0.139&  0.129\\
\hline
\end{tabular}}
\label{tab:KMbias-var.KGM}
\end{table}

\begin{table}[h]\centering
\caption{Simulation results based on the Koziol$-$Green model for
variance of the bias of the four K$-$M mean lifetime estimators
$\hat{S}^{KM}_{\mbox{mean}}$, $\tilde{S}^{KM}_{\mbox{mean}}$, $\hat{S^*}^{KM}_{\mbox{mean}}$ and $\tilde{S^*}^{KM}_{\mbox{mean}}$.} \scalebox{0.7}{
\begin{tabular}{lcccccccccc}\hline
$P_{\%}$&n=30&n=50&n=100&n=150&&n=30&n=50&n=100&n=150\\\hline
&\multicolumn{4}{c}{Variance of bias of $\hat{S}^{KM}_{\mbox{mean}}$}&&\multicolumn{4}{c}{Variance of bias of $\tilde{S}^{KM}_{\mbox{mean}}$}\\
10& 0.004& 0.002& 0.001& 0.000&& 0.010& 0.005& 0.002& 0.001\\
20& 0.008& 0.006& 0.003& 0.002&& 0.019& 0.013& 0.006& 0.004\\
30& 0.016& 0.012& 0.006& 0.004&& 0.037& 0.027& 0.014& 0.010\\
40& 0.024& 0.019& 0.012& 0.009&& 0.056& 0.045& 0.028& 0.021\\
50& 0.034& 0.028& 0.021& 0.016&& 0.082& 0.064& 0.049& 0.037\\
60& 0.041& 0.037& 0.029& 0.025&& 0.096& 0.088& 0.067& 0.058\\
70& 0.040& 0.038& 0.034& 0.032&& 0.092& 0.090& 0.081& 0.074\\
80& 0.030& 0.030& 0.029& 0.029&& 0.071& 0.074& 0.069& 0.071\\
90& 0.011& 0.011& 0.013& 0.013&& 0.034& 0.032& 0.034& 0.035\\
&\multicolumn{4}{c}{Variance of bias of $\hat{S^*}^{KM}_{\mbox{mean}}$}&&\multicolumn{4}{c}{Variance of bias of $\tilde{S^*}^{KM}_{\mbox{mean}}$}\\
10& 0.021& 0.008& 0.003& 0.001&& 0.034& 0.014& 0.004& 0.002\\
20& 0.031& 0.019& 0.007& 0.004&& 0.053& 0.032& 0.012& 0.008\\
30& 0.056& 0.035& 0.015& 0.011&& 0.095& 0.061& 0.027& 0.020\\
40& 0.078& 0.056& 0.031& 0.022&& 0.135& 0.099& 0.057& 0.039\\
50& 0.116& 0.077& 0.054& 0.039&& 0.201& 0.136& 0.098& 0.070\\
60& 0.117& 0.100& 0.073& 0.059&& 0.209& 0.181& 0.132& 0.108\\
70& 0.101& 0.092& 0.082& 0.073&& 0.183& 0.171& 0.151& 0.135\\
80& 0.063& 0.064& 0.061& 0.063&& 0.121& 0.128& 0.118& 0.123\\
90& 0.018& 0.019& 0.022& 0.023&& 0.045& 0.045& 0.049& 0.051\\\hline
\end{tabular}}
\label{tab:KM.var}
\end{table}

%
The results show that, for both estimators, the bias increases as censoring increases until a particular censoring level, then declines. That particular censoring level falls in the range 60 to 80. Above that censoring level the bias decreases as censoring increases, and decreases much more rapidly for the corrected estimator than for the K--M estimator. In addition, the bias for the corrected estimator at $P_{\%}=90$ censoring is positive for all sample sizes. This behaviour at high censoring levels does not appear in Stute and Wang (1994\nocite{stut:94b:thejack}) who investigated the bias up to only $P_{\%}=66.7$, but it is easily seen from Table~\ref{table:km.summary} that if censoring is $100\%$, then $\delta_{(n)}=0$, so the bias is 0.
A similar trend is observed for the variance of the bias of the two estimators.

We have computed also the bias of the jackknife estimate and its variance based on both the
modified estimators $\hat{S^*}^{KM}_{\mbox{mean}}$ and
$\tilde{S^*}^{KM}_{\mbox{mean}}$. The modification is based
on the predicted difference quantity approach where $\tilde{Y}_{(n)}$ is
replaced by $Y_{(n)}+\nu$ ($W_{\nu}$ in Table \ref{table:imp_methods}), as discussed in Khan and Shaw (2013b\nocite{Kha:sha:13:OnDeal}).
The bias and its variance are shown in Table \ref{tab:KMbias-var.KGM}
and \ref{tab:KM.var} respectively (the second sub-table for both tables).
The results demonstrate that under the modified approach, slightly larger bias and variance
estimates are obtained. Their overall trends are similar to those of the original estimators.

\subsection{Second Simulation Study}
In the second simulation, survival times are generated from four skewed distributions
, and censoring times independently from other specified distributions,
as listed in Table \ref{table:pdf}.
Datasets are generated randomly subject to the restriction $\delta_{(n-1)}=0$,
and, for the original jackknife formula, with the additional restriction $\delta_{(n)}=1$.

\begin{table}[h]
\caption{The failure time distributions with their corresponding censoring distributions.}
\centering
\scalebox{0.9}{
\begin{tabular}{ll}\hline
Failure time distributions&Censoring distributions\\\hline
Log-normal\,(1.1, 1):$\frac{1}{\sqrt{2\pi}}\frac{\exp(-(\log t-1.1)^2/2)}{t}$ &Uniform: U$\,(a, 2a)$\\
Exponential\,(0.2):$\frac{1}{5} \exp(-\frac{t}{5})$&Exponential: Exp$\,(\lambda)$\\
Gamma\,(4, 1):$\frac{1}{\Gamma(4)}t^3\exp(-t)$&Uniform: U$\,(a, 2a)$\\
Weibull\,(3.39, 3):$\frac{3}{38.96}t^2\exp(-\frac{t^3}{38.96})$&Uniform: U$\,(a, 2a)$\\
\hline
\end{tabular}}
\label{table:pdf}
\end{table}

In the case when $T\sim \mbox{Exp}\,(0.2)$ and $C\sim \mbox{Exp}\,(\lambda)$ for a chosen level of censoring percentage
$P_{\%}$, it follows that $Y$ and $\delta$ are independent with
$P_{\%}/100=\mbox{pr}\,(\delta=0)=\lambda/(0.2+\lambda)$.
For censoring time the Uniform distribution over
the range $[a, 2a]$ is chosen.
%
\begin{figure}
\centering \subfigure[For T $\sim$ LN\,(1.1, 1) \& C $\sim$ U$\,(a, 2a)$.]{
   \includegraphics[scale=0.50]{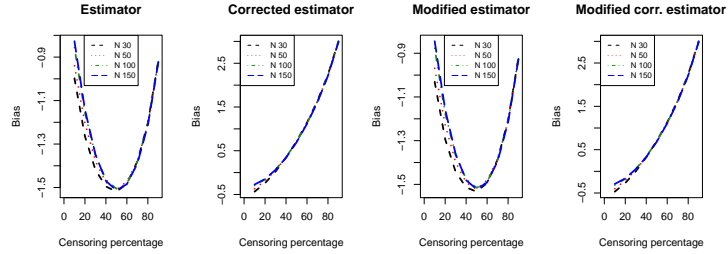}
   \label{fig:ex2ln.bias}
 }
\subfigure[For T $\sim$ EX\,(0.2) \& C $\sim$ EX$\,(\lambda)$.]{
   \includegraphics[scale=0.50]{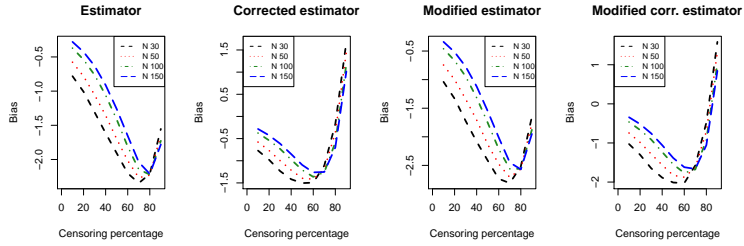}
   \label{fig:ex2ex.bias}
 }
 \subfigure[For T $\sim$ G\,(4, 1) \& C $\sim$ U$\,(a, 2a)$.]{
   \includegraphics[scale=0.50]{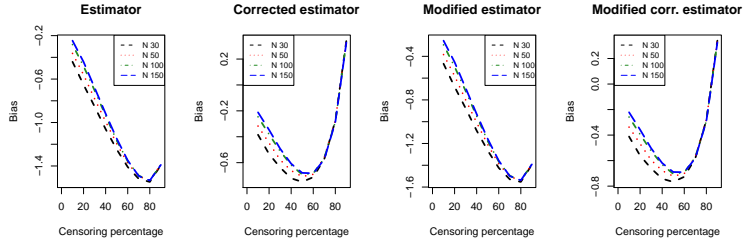}
   \label{fig:ex2ga.bias}
 }
 \subfigure[For T $\sim$ WB\,(3.39, 3) \& C $\sim$ U$\,(a, 2a)$.]{
   \includegraphics[scale=0.50]{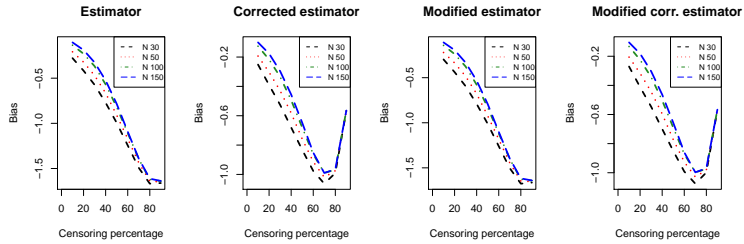}
   \label{fig:ex2wb.bias}
 }
\caption{The bias of the K--M mean lifetime estimators
$\hat{S}^{KM}_{\mbox{mean}}$, $\tilde{S}^{KM}_{\mbox{mean}}$, $\hat{S^*}^{KM}_{\mbox{mean}}$ and $\tilde{S^*}^{KM}_{\mbox{mean}}$ in 10000
simulation runs.
}
\label{fig:ex2bias}
\end{figure}

\begin{figure}
\centering \subfigure[For T $\sim$ LN\,(1.1, 1) \& C $\sim$ U$\,(a, 2a)$.]{
   \includegraphics[scale=0.50]{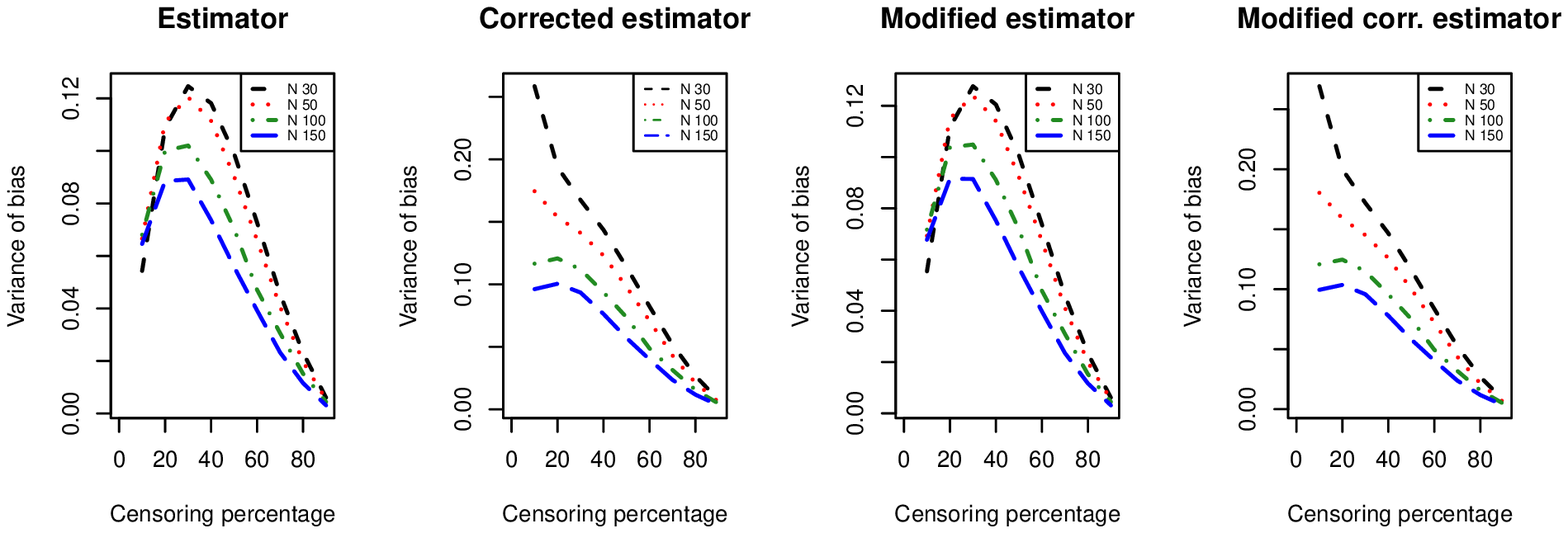}
   \label{fig:ex2ln.bias-var}
 }
\subfigure[For T $\sim$ EX\,(0.2) \& C $\sim$ EX$\,(\lambda)$.]{
   \includegraphics[scale=0.50]{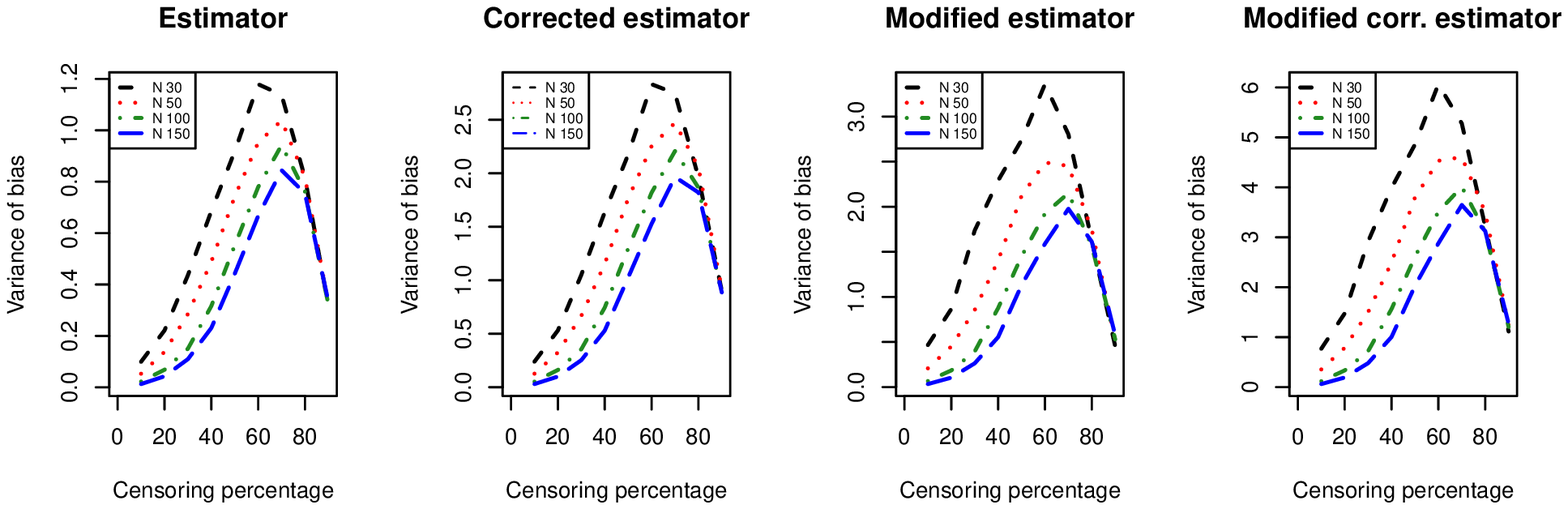}
   \label{fig:ex2ex.bias-var}
 }
 \subfigure[For T $\sim$ G\,(4, 1) \& C $\sim$ U$\,(a, 2a)$.]{
   \includegraphics[scale=0.50]{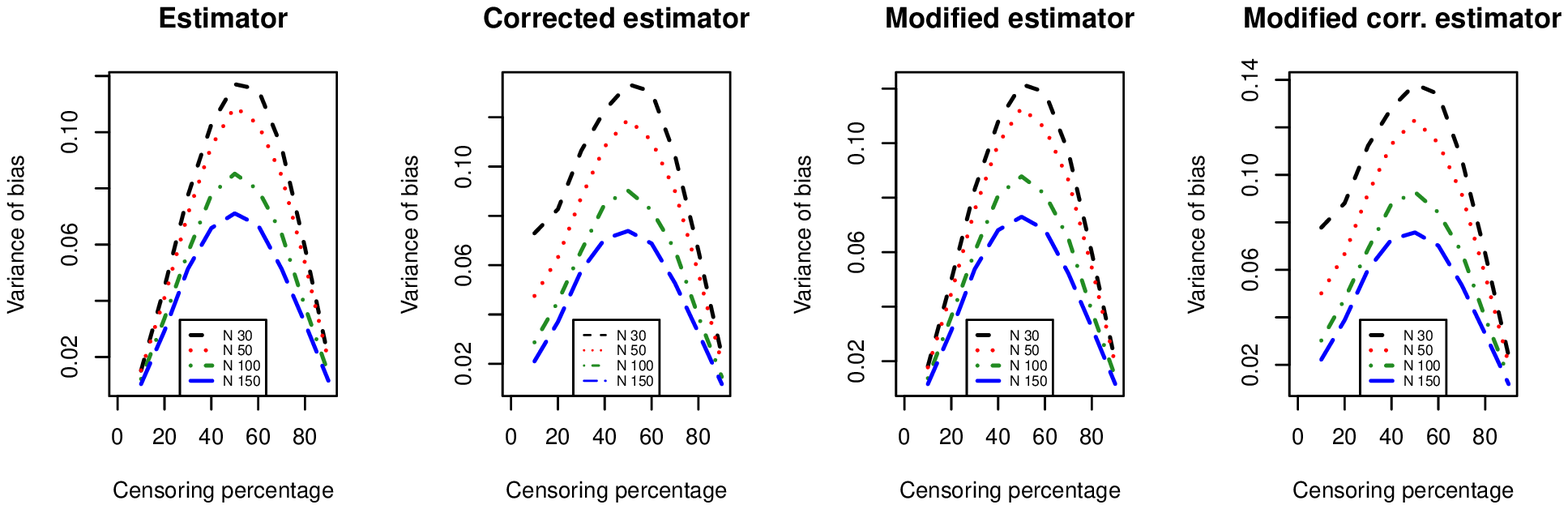}
   \label{fig:ex2ga.bias-var}
 }
 \subfigure[For T $\sim$ WB\,(3.39, 3) \& C $\sim$ U$\,(a, 2a)$.]{
   \includegraphics[scale=0.50]{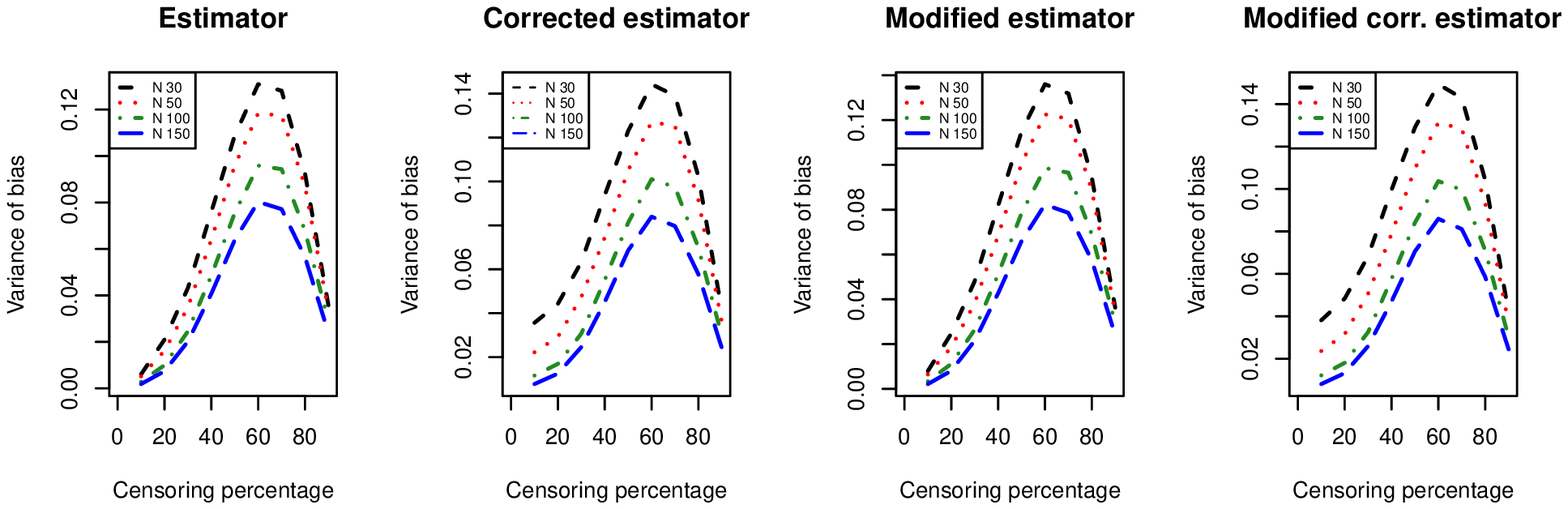}
   \label{fig:ex2wb.bias-var}
 }
\caption{The variance of the bias of the K$-$M mean lifetime estimators
$\hat{S}^{KM}_{\mbox{mean}}$, $\tilde{S}^{KM}_{\mbox{mean}}$, $\hat{S^*}^{KM}_{\mbox{mean}}$ and $\tilde{S^*}^{KM}_{\mbox{mean}}$ in 10000
simulation runs.}
\label{fig:ex2bias-var}
\end{figure}

We use four samples $n = 30,\,50,\,100,\,150$. The jackknife estimate of bias and its variance for all four estimators from $10,000$
simulated datasets are shown in Fig.~\ref{fig:ex2bias} and \ref{fig:ex2bias-var} (both shown in supplementary document) respectively.
The associated modification is here carried out using method $W_{\nu}$ of Table \ref{table:imp_methods}, described fully in (Khan and Shaw, 2013b\nocite{Kha:sha:13:OnDeal}).

Fig.~\ref{fig:ex2ln.bias}, \ref{fig:ex2wb.bias} and \ref{fig:ex2ln.bias-var}, \ref{fig:ex2wb.bias-var} reveal similar results to our large simulation based Koziol--Green model example. For example, given the modification, the bias estimate is bound to be higher. This seems to be true also for the variance estimate. In addition, we find that for both actual and modified estimators the trend in bias differs for different censoring levels, but they behave similarly under different lifetime distributions (see Fig.~\ref{fig:ex2bias}). The relationship between bias and censoring level varies substantially between the distributions and the sample sizes.
For a log-normal distribution, the bias for the estimators except for the corrected estimators tends to increase as $P_{\%}$ increases until 50. The maximum bias for the other distributions investigated occurs between 60\% and 80\% censoring. Under the Exponential lifetime distribution the bias behaves very similarly to that of the Koziol--Green proportional hazards model. Given that the estimators are original or modified the corrected estimators seem to be overestimated in the higher censoring points (i.e.,~the bias becomes positive in higher censoring).

The variance (Fig.~\ref{fig:ex2bias-var}) of bias for estimators also differs according to sample sizes and censoring level. The variance generally reaches a maximum at some censoring level between $50\%$ and $70\%$, then declines. However, for the corrected estimators under a log-normal distribution the variance decreases consistently as censoring increases (see Fig.~\ref{fig:ex2ln.bias-var}).

\subsection{Third Simulation Study}
\begin{figure}[ht]
\centering
\centering   \subfigure[Bias]{
   \includegraphics[scale=0.5]{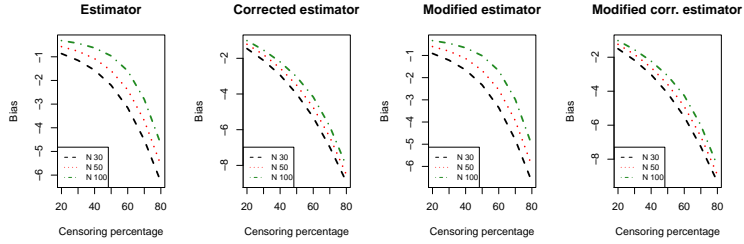}
   \label{fig:ex3ln.bias}
 }
 \subfigure[Variance of bias]{
   \includegraphics[scale=0.5]{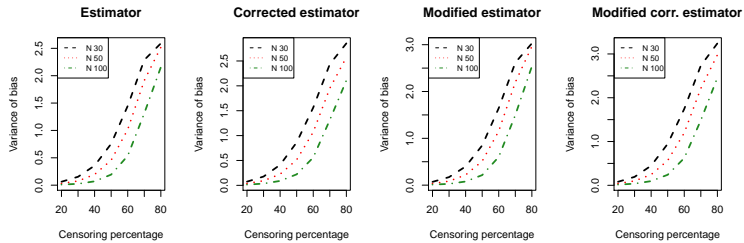}
   \label{fig:ex3ln.bias-var}
 }
\caption{Simulation results for the third simulated example for all four K$-$M mean lifetime estimators $\hat{S}^{KM}_{\mbox{mean}}$, $\tilde{S}^{KM}_{\mbox{mean}}$, $\hat{S^*}^{KM}_{\mbox{mean}}$ and $\tilde{S^*}^{KM}_{\mbox{mean}}$ under the log-normal AFT model at different censoring points. Lowess smooths are superimposed.}
\label{fig:kmbias.ex3}
\end{figure}

This simulation study is conducted to investigate how the modified estimators behave relative to the
original estimators when lifetimes are modeled as an AFT model that has the form
\begin{equation}
Z_i=\alpha+X_i^T\beta +\sigma \varepsilon_i,~~i=1,\cdots,n\qquad \varepsilon_i\sim N(0,1)~\text{for}~i=1,\cdots,n
\label{eq:LNaft}
\end{equation}
where $Z_i=\log\,(T_i)$, $\mathbf{X}$ is the covariate vector,
$\alpha$ is the intercept term, $\beta$ is the unknown $p\times 1$
vector of true regression coefficients.
The logarithm of the true survival time
is generated from the true model (\ref{eq:LNaft}).
The logarithm of censoring time is assumed to be distributed as U$(a, 2a)$ where $a$ is chosen analytically in the same way as done in the previous example. We consider five covariates $\mathbf{X}=\,(X_1,\,X_2,\,X_3,\,X_4,\,X_5)$ each of which is generated using U$(0,1)$, seven $P_{\%}$ points, and three samples $n=30$, $50$ and $100$. The coefficients of the covariates are chosen as $\beta_j=j+1$ where $j=1,\cdots,5$ and $\sigma=1$. Of the five proposed imputation approaches of Table~\ref{table:imp_methods} and Khan and Shaw (2013b\nocite{Kha:sha:13:OnDeal}), the resampling based conditional mean approach ($W_{\tau^{\ast}_{m}}$) is found to have the least bias, and the results for $W_{\tau^{\ast}_{m}}$ from $10,000$ simulation runs are shown in Fig.~(\ref{fig:kmbias.ex3}).

\section{Discussion}

The behavior of bias for the K--M lifetime estimators is influenced by many factors in practice. For example, the nature of the distributions to be used for lifetimes, the censoring rate, the sample size, whether the lifetimes are modeled with the covariates and so on. To explore the behaviour of the jackknife bias for K--M estimators under various conditions (in particular, censoring levels) a large simulation is required.
Our simulation studies go beyond the small simulation study in Stute and Wang (1994\nocite{stut:94b:thejack}) and show clear differences from many of their results. In particular, the bias (Eq.~(\ref{eq:jkbias}) and (\ref{eq:jkbias*})) will be 0 at $0\%$ censoring and increases as the censoring level increases. However, the bias will also tend to 0 as the censoring level tends to $100\%$ (because the bias is 0 when either $\delta_{(n-1)}$ or $\delta_{(n)}$ is 0). Therefore, as shown in the figures, the bias increases up to a particular censoring level (typically $50\%-80\%$) but then reduces. The variance of the bias shows similar behaviour. Note also that the bias for the corrected estimators tends to be overestimated at the higher censoring level ($90\%$).

We propose the modified K--M survival estimator, the modified jackknife estimate of bias for K--M estimator and the modified bias corrected K--M estimator. The modification allows one pair of observations ($\delta_{(n)}=0$, $\delta_{(n-1)}=0$) to contribute to the bias calculation. So our modifications reduce the original conditions needed for jackknife estimation of bias ($\delta_{(n-1)}=0$, $\delta_{(n)}=1$) to the single condition $\delta_{(n-1)}=0$.
The modified jackknife estimate also prevents the K--M estimator from being badly underestimated by the jackknife estimate when the largest observation is censored. For calculating bias and its variance with the proposed and existing jackknifing procedures we have provided a publicly available package \emph{jackknifeKME} (Khan and Shaw, 2013a\nocite{has:ewa:Rpack:jackknifeKME}) implemented in the R programming system.

\section{Acknowledgements}
The first author is grateful to the Centre for Research in Statistical Methodology (CRiSM), Department of Statistics, University of Warwick, UK for offering research funding for his PhD study.

\bibliography{b2ndyear}
\bibliographystyle{model2-names}

\end{document}